\documentclass[fleqn,usenatbib]{mnras}

\usepackage{newtxtext,newtxmath}

\usepackage[T1]{fontenc}

\DeclareRobustCommand{\VAN}[3]{#2}
\let\VANthebibliography\thebibliography
\def\thebibliography{\DeclareRobustCommand{\VAN}[3]{##3}\VANthebibliography}

\usepackage{graphicx}	
\usepackage{amsmath}	

\newcommand{\Msun}{M$_{\odot}$}

\title[Tertiaries to ELM WD binaries]{Most extremely low mass white dwarfs with non-degenerate companions are inner binaries of hierarchical triples}

\author[F. Lagos et al.]{
Felipe Lagos-Vilches,$^{1}$\thanks{E-mail: felipe.lagos.vilches@gmail.com}
Mercedes Hernandez,$^{2}$
Matthias R. Schreiber,$^{2}$
Steven G. Parsons,$^{3}$
\newauthor
Boris T. G\"ansicke$^{1,4}$
\\
$^{1}$ Department of physics, University of Warwick, Gibbet Hill, Coventry CV4~7AL, UK\\
$^{2}$ Departamento de F\'isica, Universidad T\'ecnica Federico Santa Mar\'ia, Avenida Espa\~na 1680, Valpara\'iso, Chile\\
$^{3}$ Department of Physics and Astronomy, University of Sheffield, Sheffield S3 7RH, UK\\
$^{4}$  Centre for Exoplanets and Habitability, University of Warwick, Coventry CV4 7AL, UK\\
}

\date{Accepted XXX. Received YYY; in original form ZZZ}

\pubyear{2015}

\begin{document}
\label{firstpage}
\pagerange{\pageref{firstpage}--\pageref{lastpage}}
\maketitle

\begin{abstract}

Extremely-low-mass white dwarfs (ELM WDs) with non-degenerate companions are believed to originate from solar-type main-sequence binaries undergoing stable Roche lobe overflow mass transfer when the ELM WD progenitor is at (or just past) the termination of the main-sequence. This implies that the orbital period of the binary at the onset of the first mass transfer phase must have been $\lesssim 3-5$\,d. This prediction in turn suggests that most of these binaries should have tertiary companions since $\approx 90$ per cent of solar-type main-sequence binaries in that period range are inner binaries of hierarchical triples. Until recently, only precursors of this type of binaries have been observed in the form of EL\,CVn binaries, which are also known for having tertiary companions. Here, we present high-angular-resolution images of TYC\,6992-827-1, an ELM WD with a sub-giant (SG) companion, confirming the presence of a tertiary companion. Furthermore, we show that TYC\,6992-827-1, along with its sibling TYC\,8394-1331-1 (whose triple companion was detected via radial velocity variations), are in fact descendants of EL\,CVn binaries. Both TYC\,6992-827-1 and TYC\,8394-1331-1 will evolve through a common envelope phase, which depending on the ejection efficiency of the envelope, might lead to a single WD or a tight double WD binary, which would likely merge into a WD within a few Gyr due to gravitational wave emission. The former triple configuration will be reduced to a wide binary composed of a WD (the merger product) and the current tertiary companion.

\end{abstract}




\section{Introduction}
Current theories of close white dwarf (WD) binary formation and evolution are rather limited. The most widely used prescriptions for mass transfer interactions and
angular momentum loss are relatively simple conservation equations containing a number of neither theoretically nor observationally well constrained parameters \citep[e.g., ][]{Nelemans_Tout2005,Zorotovic2010,Ivanova2013,Belloni_Schreiber2023}. Therefore, binary population simulations are unable to reliably predict detailed characteristics of WD binary populations \citep{Nelemans2001}. Even worse, the existence of some individual systems 
appears to be paradoxical \citep[e.g., ][]{Bours_2015} and hundreds of WDs in wide WD+WD binaries have been found with discrepant cooling ages \citep{Heintz2022, Heintz2023}. Consequently, observational constraints on models for the formation and evolution of close WD binaries are urgently required. 

Recently, \citet{Lagos_2020} showed that the occurrence rate of tertiary companions to close WD binary stars can provide information on the initial binary separation, thereby providing key insights into the formation of the inner binary. EL\,CVn binaries are eclipsing systems consisting of an extremely low mass (ELM) helium WD precursor (extremely low mass is defined here as less than $0.3\, \mathrm{M_{\odot}}$) with an A/F -type main sequence companion \citep{maxtedetal14-1}. According to \citet{Chen_2017}, these pre-ELM WDs with A/F main sequence companions form from solar-type close main sequence binaries with periods shorter than a few days through stable but non-conservative mass transfer. The short initial period is required to produce early mass transfer when the core of the progenitor of the pre-WD still has the low mass that we observe today in EL\,CVn binaries. 

By observing a sample of solar-type spectroscopy binaries, \citet{Tokovinin06} found a negative correlation between the orbital period and the frequency of tertiary companions, going from $34$ to $96$ per cent for orbital periods $P>12$ and $P<3$ days respectively (after correcting for incompleteness). This trend has been further confirmed by \citet{Laos2020} and is supported by the fact that many binaries with orbital periods of a few days or less have tertiary companions \citep{Pribulla2006,Raghavan2010,Tokovinin2014,Hwang2023}. This correlation implies that the vast majority of EL\,CVn binaries should host a distant tertiary. Indeed, high contrast imaging of nearby EL\,CVn binaries confirmed that most of them are inner binaries of hierarchical triples \citep{Lagos_2020}, and new EL\,CVn binaries with tertiary companions continue to be discovered \citep{Lee2024}.

While \citet{Lagos_2020} only considered EL\,CVn binaries, ELM WDs should also form when mass transfer is initiated in a very close binary star. However, the companion stars of the large sample of ELM WDs in the ELM survey \citep{Brown_2010,Kosakowski_2020,Brown_2020} are typically WDs. This implies that two evolutionary sequences are possible. Either the ELM WD formed first, which would require stable mass transfer to occur in a close main sequence binary (such as during the formation of EL\,CVn binaries) and the WD companion forms in the second mass transfer phase (most likely common envelope evolution), or the first phase of mass transfer produces a short period WD plus main sequence binary and the ELM WD forms in the second (stable) mass transfer phase \citep[e.g. ][]{Li_2023}. 

Recently, \citet{Parsons_2023} presented three low-mass WDs with sub-giant (SG) companion stars. These objects are perfectly consistent to have formed through dynamically stable mass transfer. The smaller the core mass of the initially more massive star at the onset of mass transfer, the shorter the initial period must have been, and the smaller is the resulting WD mass. 

Two of those binaries contain ELM WDs, which implies that the period of the main-sequence binary progenitor must have been short, suggesting these systems should host distant tertiaries. Remarkably, \citet{Parsons_2023} confirmed through radial velocity variations that one of the ELM WD+SG binaries, TYC 8394-1331-1 (hereafter TYC\,8394), indeed has a tertiary companion. \citet{Parsons_2023} therefore predicted that TYC\,6992-827-1 (hereafter TYC\,6992), the second ELM WD+SG binary, should also be a triple system (the stellar parameter of TYC\,6992 are available in Table\,\ref{tab:WD+SG params}). 

Here we present high-angular resolution VLT/SPHERE observations of TYC\,6992 and indeed find a tertiary companion, exactly as predicted. We complement these observational results with numerical simulations of stable mass transfer using the Modules for Experiments in Stellar Astrophysics (MESA) code \citep[][]{Paxton2011,Paxton2013,Paxton15,Paxton2018,Paxton2019,Jermyn2023}. Combining our findings with those of \citet{Tokovinin06}, we estimate that $\gtrsim70$ percent of ELM WDs with non-degenerate companions should be the inner binaries of hierarchical triples.

\begin{table}
 
	\caption{Stellar and orbital parameters for TYC\,6992-827-1 reported by \citet{Parsons_2023}. ELM WD parameters were obtained from HST spectral fitting using the \citet{Althaus2013} cooling tracks, while sub-giant parameters were acquired through UVES spectral fitting. The orbital parameters include observations from FEROS, UVES, Du Pont, and CHIRON.}
	\resizebox{\columnwidth}{!}{\begin{tabular}{lllcc}
	\hline
	Stellar parameter &  ELM WD & Sub-Giant& \\
	\hline
	Mass [$\mathrm{M_{\odot}}$]            & $0.28\pm0.01$     &$1.31 \pm 0.14$ \\
    Radius [$\mathrm{R_{\odot}}$]            & $0.0235\pm0.001$     &$3.45 \pm 0.12$ \\
    log $g$ [dex]  & $7.14 \pm 0.02$ &  $3.48 \pm 0.04$\\
   $\mathrm{T_{eff}}$ [K]    &   $15750 \pm 50$   & $5250 \pm 50$\\
    Cooling time [Myr]        & $6.9 \pm 0.1$     & -\\
  \hline
	&Orbital parameters&  \\
  \hline
	Orbital period [d]        & $41.45 \pm 0.01$ \\
    Semi-major axis [au]      &$0.27   \pm 0.01$ \\
    Eccentricity              & $0.013 \pm 0.06$ \\	
	\hline
	\end{tabular}}
	\label{tab:WD+SG params}
\end{table}

\section{Observations and characterization of the tertiary companion}

Following an observational strategy similar to that described in \cite{Lagos_2020}, we obtained high contrast images of TYC\,6992 (program ID 110.243S.001) with the VLT/SPHERE instrument using the InfraRed Dual-band Imager and Spectrograph
 \citep[IRDIS,][]{Dohlen_2008} in the dual band imaging mode \citep{Vigan_2010}. Coronagraphic images were acquired using the $H2$ ($\lambda_{H2}=1593$\, nm) and $H3$ ($\lambda_{H3}=1667$\,nm) filter and the N-ALC-YJH-S coronagraph. The IRDIS data were pre-processed (dark background subtraction, flat-fielding, bad-pixels correction and frame recentring) with the \textsc{sphere} python package\footnote{\url{https://github.com/avigan/SPHERE}} version 1.6.1 \citep{vlt-sphere}.

\begin{figure}
\includegraphics[width=\columnwidth]{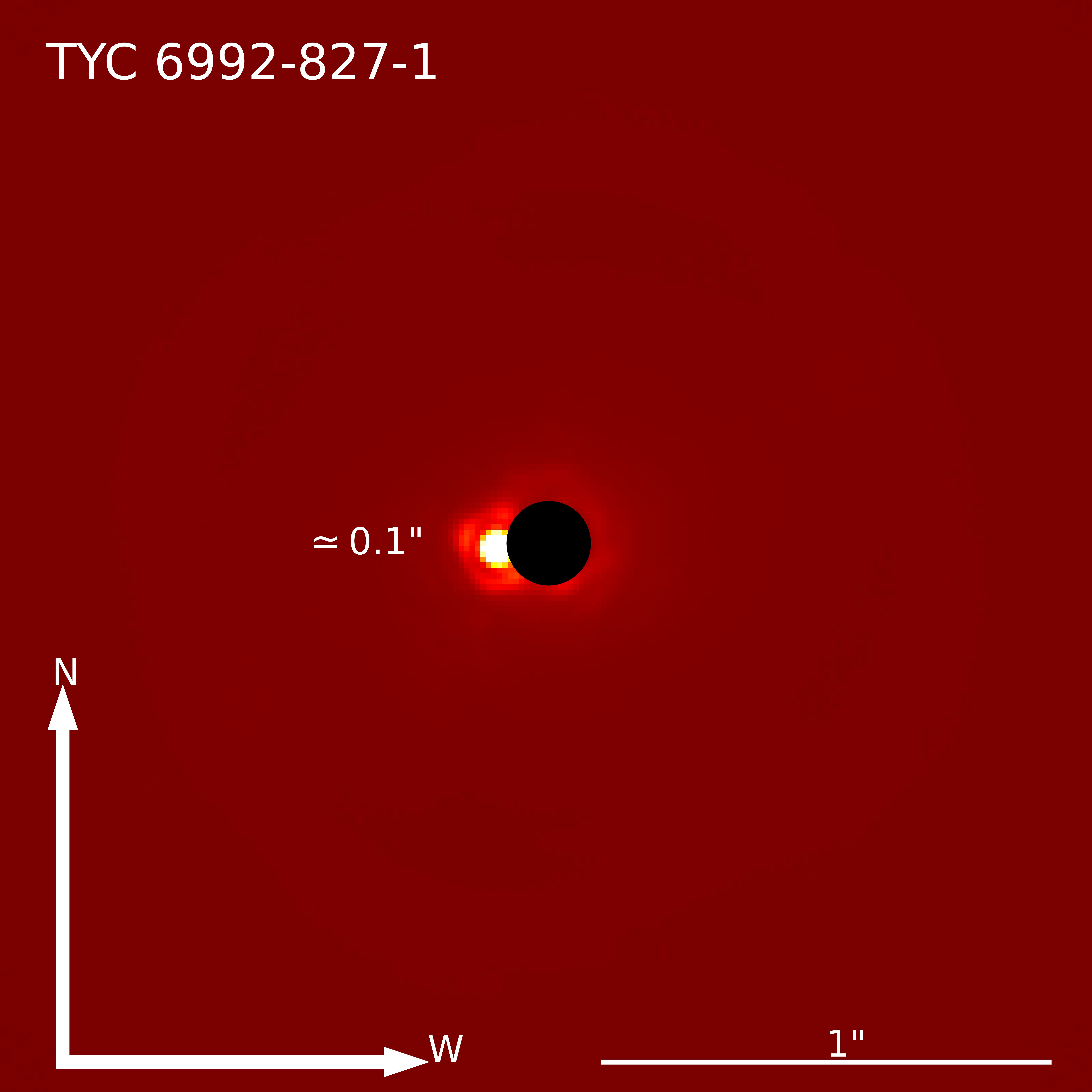}
\caption{SPHERE/IRDIS images of TYC\,6992 in the H2 band.
The black circle denotes the position of the ALC\_YJH\_S coronograph ( RA=00:01:13.40, DEC=-33:27:06.84). The projected separation of the tertiary companion to the central binary is $117$ mas corresponding to $\sim59$\,au at the distance of the system.}
    \label{fig:SPHERE_TYC_6992}
\end{figure}

The coronagraphic frame shown in Fig. \ref{fig:SPHERE_TYC_6992} clearly reveals the presence of a tertiary candidate located at an angular separation of $\approx 117$\,mas from the ELM WD + SG inner binary, which corresponds to a projected separation of $\approx 59$\,au. For each of the 112 (7) coronagraphic (flux calibration) frames we performed aperture photometry to the tertiary candidate using an aperture of 4 pixels. This aperture is chosen to minimize the observed overlap between the tertiary and the ELM WD + SG binary (Fig.\,\ref{fig:SPHERE_TYC_6992}). Contamination coming from the quasi-azimuthal speckle patterns produced by the coronagraph at the location of the detection was estimated by calculating the flux enclosed within one aperture at the reflected location of the detection relative to the SG-WD binary. By averaging the photometry obtained for the 112 (7) coronagraphic (flux calibration) frames we derived a tertiary-to-inner binary flux ratio of $4.3\pm0.1$ and $ 4.1\pm0.2$ per cent in the $H2$ and $H3$ filters, respectively. Given that the WD flux is negligible in these filters relative to the SG, we can translate these flux ratios into magnitudes through synthetic photometry to the best-fit model spectrum of the SG star calculated by \cite{Parsons_2023}. By doing so, we derived $H2\approx H3\approx 12.72$ for the tertiary candidate.

If the tertiary is still on the main sequence, the derived magnitudes are consistent with a star of spectral type K or M, since it must have been less massive than the SG progenitor, most likely a G/late-F main-sequence star. In fact, by inspecting the isochrones for main-sequence low-mass stars of \citet{Baraffe_2015} and assuming a current age of $9$\,Gyr for the entire system ($\approx7$ Myr of WD cooling time from Table \ref{tab:WD+SG params} plus $8$ Gyr before the start of mass transfer according to MESA evolution tracks for a $1.1\,\mathrm{M_{\odot}}$ ELM WD progenitor), we found that the calculated $H2$ and $H3$ magnitudes are roughly consistent with a tertiary of $0.7-0.8$\,$\mathrm{M_{\odot}}$.

The probability $P_\mathrm{bkg}$ of the tertiary candidate being a false positive detection (i.e. not gravitationally bound to the ELM WD + SG binary) can be estimated by modeling as a Poisson process the occurrence of a background source down to a limiting magnitude $H_\mathrm{lim}=13$ within a radius of $\Theta=117$\,mas from the SG-WD binary. If $\rho(H_\mathrm{lim})$ is the surface density of background sources down to the limiting magnitude $H_\mathrm{lim}$, then

\begin{equation}
  P_\mathrm{bkg}(\Theta,m_\mathrm{lim})= 1-e^{-\pi \Theta^2 \rho(H_\mathrm{lim})}. 
\end{equation}  

To estimate $\rho(H_\mathrm{lim})$, we use the Besan\c{c}on galaxy model\footnote{\url{https://model.obs-besancon.fr/}} \citep{Robin_2004} to generate a synthetic 2MASS photometric catalogue of point sources within one square degree and magnitude $H<H_\mathrm{lim}$ centered on the coordinates of TYC\,6992. We found that the probability of the companion to be a background source is $0.001$ per cent, suggesting that the triple candidate is most likely member of TYC\,6992.

\section{Formation scenario of ELM WD+SG binaries}

WDs with masses $\lesssim0.48\,\mathrm{M_{\odot}}$, such as the one observed in TYC 6992 (and TYC 8394), cannot be formed via single stellar evolution within the Hubble time. Instead, they are the product of a mass transfer event that truncates the evolution of the WD progenitor \citep{rebassa-mansergasetal-2011}. Apparently single low-mass WDs likely formed through merger events \citep{schreiberetal16-1,zorotovic+schreiber17-1}. 

Depending on the response of the donor star to mass loss, which depends largely on the mass ratio, mass transfer can proceed in an unstable and runaway fashion, triggering a common envelope (CE) event, or in a much more stable and self-regulated manner \citep[see][for recent reviews]{Ivanova2013,Belloni_Schreiber2023}. 
Given its complex magneto-hydrodynamic nature, simulations of CE evolution capturing all the required details are currently unavailable. To nevertheless predict the outcome of CE evolution instead the so-called energy or $\alpha$ formalism is frequently used \citep[e.g. ][]{Webbink_1984,Zorotovic2010,zorotovic_2014}. In this framework simple energy conservation arguments allow to estimate the final orbital period and stellar masses of a binary system at the end of the CE phase for a given value of the CE efficiency parameter. Additionally, the $\alpha$ formalism allows to reconstruct possible evolutionary scenarios and estimating the binary period and stellar masses at the onset of the common envelope for a given measured post-CE configuration \citep{Nelemans_1998,Zorotovic2010}. For stable mass transfer, performing detailed simulations, e.g. with MESA, is required. Such simulations allow to realistically track the evolution of both the stellar and orbital parameters.

In what follows, we first show that CE evolution fails at reproducing the observed properties of the ELM WD + SG binaries. We then investigate the feasibility of the stable mass transfer scenario through a set of MESA simulations.

\subsection{Excluding common envelope evolution}

\citet{Parsons_2023} already pointed out that CE evolution is unlikely to explain the formation of both ELM WD + SG binaries characterized in their paper. 
We here further confirm this suggestion following the formalism proposed by \citet[][their appendix A]{Chen_2017}. In short, using the energy formalism to model the CE phase, the complete ejection of the envelope and the formation of a post-CE binary are only possible if the following criterion is satisfied:

\begin{equation}
    \frac{0.462\alpha\lambda}{2(q_\mathrm{i}-q_\mathrm{f})} \left[ \frac{q_\mathrm{f}}{q_\mathrm{i}} \left( \frac{1+q_\mathrm{i}}{1+q_\mathrm{f}} \right)^{1/3} \left( \frac{P_\mathrm{i}}{P_\mathrm{f}} \right)^{2/3} -1 \right] \left( \frac{q_\mathrm{i}}{1+q_\mathrm{i}} \right)^{1/3} - 1 \geq 0.
    \label{eq:CE criterion}
\end{equation}

Here $q_\mathrm{i}$ ($q_\mathrm{f}$) is the initial (final) mass ratio $M_\mathrm{d}/ \mathrm{M_a}$ between the donor and the accretor star, $P_\mathrm{i}$($P_\mathrm{f}$) is the orbital period at the onset (termination) of the CE phase, $\alpha$ the CE efficiency and $\lambda$ the envelope-structure parameter (often called binding energy parameter). To obtain $P_\mathrm{i}$, we assumed that the radius of the ELM WD progenitor (estimated with the radius-core mass relation for giants of \citet{Rappaport_1995} ) is equal to its Roche radius (estimated with the Roche lobe radius equation of \citet{Paczynski_1971}) at the onset of CE evolution. Then, by using Kepler's third law, the orbital separation at the onset of CE is translated into orbital period (Eq. A7 of \citet{Chen_2017}). Since the ELM WD progenitor (the donor) must have been more massive than its companion, we have $q_\mathrm{i}>1$. Therefore, we explore values of $q_\mathrm{i}$ between $1$ and $3$. To ensure the most favorable scenario for a successful envelope ejection, we set $\alpha=1$ (i.e. the orbital energy is fully converted into work in the expanding envelope) and $\lambda=2$ (i.e. the envelope is assumed to be loosely bound to the core, as might be expected during the final stages of post-main sequence evolution \citep{Claeys_2014})

For TYC\,6992, the condition given by Eq. \ref{eq:CE criterion} is only met for WD masses $\gtrsim 0.56\,\mathrm{M_{\odot}}$ (see Fig.\,\ref{eq:CE criterion}), far above the $0.28\,\mathrm{M_{\odot}}$ of the ELM WD. Unsurprisingly, we arrive at the same result for TYC\,8394, whose ELM WD has a mass of $0.24\,\mathrm{M_{\odot}}$. Therefore, we confirm that CE evolution fails to explain their current configuration.

\subsection{Evolution through stable mass transfer}
\label{sec:stable mass transfer}
We next investigated whether the stable mass transfer scenario is able to reproduce the configuration of the ELM WD + SG binaries. To that end, we used version r\,22.11.1 of MESA. 
The MESA equation of state is a blend of the OPAL \citep{Rogers2002}, SCVH \citep{Saumon1995}, FreeEOS \citep{Irwin12}, HELM \citep{Timmes2000}, PC \citep{Potekhin2010} and Skye \citep{Jermyn2021} equations of state. Nuclear reaction rates are a combination of rates from NACRE \citep{Angulo1999}, JINA REACLIB \citep{Cyburt2010}, plus additional tabulated weak reaction rates \citep{Fuller1985,Oda1994,Langanke2000}. Screening is included via the prescription of \citet{Chugunov2007} and thermal neutrino loss rates are from \citet{Itoh1996}. Electron conduction opacities are from \citet{Cassisi2007} and radiative opacities are primarily from OPAL \citep{Iglesias1993,Iglesias1996}, with high-temperature Compton-scattering dominated regime calculated using the equations of \citet{Buchler1976}. 

In order to ensure stable mass transfer, the initial masses of main sequence stars were chosen with consideration of the adiabatic response of the donor. This condition is typically satisfied when the initial mass ratio $q_i$ is below a critical value known as $q_\mathrm{crit}$. For cases involving conservative mass transfer and donors ranging between 1 and $6\,\mathrm{M_\odot}$, the estimated $q_{\mathrm{crit}}$ at the end of the red giant branch phase is approximately between 0.7 and 1 \citep{Ge2020}. The corresponding MESA simulation was initiated with a binary system on the zero-age main sequence, where the initial stellar masses are $1.1$ and $0.95\,\mathrm{M_\odot}$, with an initial orbital period of 4.55\,d. We assumed a mass loss fraction from the system during stable mass transfer of $\beta = 50$~per cent. Figure\,\ref{fig:mesa-tyc} illustrates the obtained formation of an extremely low mass WD through stable mass transfer. 

The track shows the evolution from the moment the main-sequence ELM WD progenitor fills its Roche lobe leading to stable and non-conservative mass transfer at a rate of approximately $10^{-7}\,\mathrm{M_{\odot}}\mathrm{yr}^{-1}$. Roche-lobe overflow ends when the star reaches a radius of around $0.72$\,$\mathrm{R}_{\odot}$. Upon Roche lobe detachment, the pre-ELM WD is formed. This pre-ELM WD reaches its first temperature peak of around $35\,000$\,K, which marks the start of the cooling phase. When the temperatures cooled down to $\simeq22\,000$\,K the first hydrogen flash occurs which causes a small loop in the HR-diagram. Subsequently, the energy released during the flash results in a sharp temperature gradient near the point of maximum energy production causing the star to expand again. Following the second maximum radius, the star evolves towards high surface temperatures at a nearly constant luminosity before transitioning to the final cooling track. At an effective temperature of $\simeq30\,000$\,K a subflash occurs which produces a similar loop as the first one but does not generate a third expansion of the star. \citet{Istrate16} presented similar evolutionary trajectories for forming ELM WDs with neutron star companions, providing a comprehensive description of the WD's evolution (see their section 2.2 for more details).

In Fig.\,\ref{fig:mesa-tyc} we also highlight the location of TYC\,6992 with a cyan star symbol. The measured parameters of the ELM WD in this system are consistent with the WD having reached the final cooling track. The very similar system, TYC\,8394 (its location is indicated by the cyan triangle), could represent a slightly earlier evolutionary stage just after the first hydrogen flash. Moreover, the 14 WD progenitors with stripped low-mass giants (yellow dots) documented by \citet{El-badry2022} are accurately reproduced by our track during the initial stable (non-conservative) mass transfer phase. 
The five EL\,CVn with tertiary companions reported by \citet{Lagos_2020} represent a later evolutionary stage which is reached after mass transfer ended and hydrogen flashes cause the star to expand again. 
We thus conclude that parameters of ELM WDs with non-degenerate companions, such as those characterized by \citet{Parsons_2023}, are consistent with having formed through stable mass transfer.

\subsection{The effect of triple dynamics prior and after the mass transfer phase}

Inner binaries in triple systems can undergo orbital shrinkage due to the combined effect of high eccentricity incursions induced by the tertiary companion through the Von Zeipel-Lidov-Kozai (ZLK) mechanism and tidal friction at (or near) periastron \citep[aka Kozai cicles with tidal friction or KCTF, ][]{Fabrycky_Tremaine2007,Shappee_Thompson2013,Naoz_Fabrycky2014}.

To estimate whether the the main sequence binary progenitor assumed in section \ref{sec:stable mass transfer} could have experienced KCTF, we simulate the evolution of a triple systems composed of an inner circular binary with stellar masses $1.1$ and $0.95\,\mathrm{M_{\odot}}$, with its semi-major axis ranging from $0.1$ to $1.1$\,au in steps of $0.2$\,au, outer binary with eccentricity $0.5$, semi-major axis of $10$\,au and tertiary mass of $0.8\mathrm{M_{\odot}}$. The mutual inclination between the inner and outer orbits is set to $70$ degrees. The choice of mutual inclination and outer eccentricity values, though somewhat arbitrary, favors the occurrence of high eccentricity incursions while remaining within the observed range derived by \citet{Borkovits2015,Borkovits2016} and \citet{Czavalinga2023}, where higher values are less common. Similarly, the outer semi-major axis is chosen to enhance the ZLK mechanism while maintaining the secular dynamical stability of the system.  

We use the \textsc{Multiple Stellar Evolution\footnote{\url{https://github.com/hamers/mse}}} \citep[MSE,][]{Hamers2021} code (version 0.87) to calculate the secular orbital evolution of the inner binary, including the effects of stellar and tidal evolution, general relativity, N-body dynamics and binary interactions. In all our simulations, we found that while the ZLK mechanism can induce inner eccentricities as high as $\approx 0.8-0.9$ during the main sequence, the tidal forces are not strong enough to cause significant inward migration. Only once the $1.1\,\mathrm{M_{\odot}}$ star reaches the red giant phase does the inner orbit start to shrink. At this stage, however, the core mass exceeds $0.28\,\mathrm{M_{\odot}}$ (i.e. the mass of the observed ELM WD), which fails to explain the current configuration of TYC\,6992. Although we cannot rule out the possibility that a different initial orbital configuration might produce more extreme eccentricity incursions (enhancing the effect of tides at periastron), such a scenario would require higher and less likely values for the outer eccentricity and mutual inclination. This, in turn, might require an initial outer semi-major axis greater than $10$\,au to ensure the secular stability of the system, which would also weaken the ZLK mechanism. We emphasize that although this analysis is based on a set of initial conditions favouring the occurrence of high eccentricity incursions, a full exploration of the orbital parameter space—beyond the scope of this paper— is needed to properly assess the role of triple dynamics in TYC\,6992.

It is important to note that, whether the inner binary was born with a short orbital period or underwent migration via KCTF, does not change the fact that the currently observed parameters of TYC\,6992 are fully consistent with a formation through stable mass transfer. The main aim of this paper is to explore the relation between ELM WDs with close non-degenerate companions and the existence of tertiaries. Whether triple dynamics played a role in the formation of the short period of the inner binary prior to mass transfer is of minor importance here. However, we note that the observed relation between the period of main sequence binaries and the existence of a tertiary \citep{Tokovinin06} cannot be fully explained by triple dynamics \citep{Borkovits2015,Borkovits2016,Moe_Kratter2018}.

In its current configuration, assuming an outer semi-major axis of $60$\,au and an eccentricity of $0.5$, the ZLK eccentricity oscillation timescale is $\approx 800$\,Myr \citep[using Eq. 41 of ][ with $m_2 \neq 0$]{Antognini2015}. As discussed in \citet{Parsons_2023}, this system is expected to undergo a common envelope phase in $\sim200-300$\,Myr, making it very unlikely that triple dynamics have any significant impact on the future evolution of TYC\,6992.

\section{Predicting the triple fraction}

Having confirmed that observed ELM WDs with non-degenerate companions must form through stable mass transfer, we now use the statistics of tertiary companions to close main sequence binaries \citep{Tokovinin06} to estimate the fraction of such binaries that should host a distant tertiary. To that end, we performed a series of 58 MESA simulations for a zero-age-main-sequence binary with primary and secondary star masses of 1.1 and 0.95\,$\mathrm{M}_{\odot}$, respectively. Again, the simulations incorporate a mass loss fraction from the system of $\beta = 50$ per cent during stable mass transfer. The initial orbital period was varied from 1 to 30\,d. 

In Fig. \ref{fig:mesa-tokl}, we present the relation between the helium core mass (i.e. the ELM WD mass) obtained from the evolution of the primary component of the binary as a function of its initial orbital period. As expected, longer initial periods produce more massive WDs as the primary is able to evolve in isolation for a longer period of time before filling its Roche lobe. 
For an initial period of $\approx 4.5$\,d, the resulting final orbital period and ELM WD mass are in agreement with those values of TYC\,6992. As is well known, a relatively tight relation exists between the final period and the WD mass of post-stable mass transfer binaries \citep[e.g.][]{Parsons_2023}. 
The period of TYC\,8394 ($51.8$\,d) is only slightly longer than TYC\,6992 ($41.5$\,d) and the WD mass is also very similar ($0.24 $ and $0.28\,\mathrm{M_{\odot}}$ for TYC\,8394 and TYC\,6992, respectively), so that a perfect match for TYC\,8394 could certainly be obtained by varying the initial masses.  

We compare the resulting relationship between initial period and WD mass with the frequency of tertiary companions as a function of the initial orbital period of a close main sequence binary from \citet{Tokovinin06}. In all cases where binaries have an initial orbital period below six days, they eventually form a WD with a mass of up to $0.29\,\mathrm{M_\odot}$. These binaries have a probability exceeding 70 percent of hosting a tertiary companion consistent with our observations.

For completeness we note that, as we fixed the initial masses, the relation between initial period and WD mass shown in Fig.\,\ref{fig:mesa-tokl} does not reflect the full set of possible solutions. However, for slightly different initial masses, that is, for binaries with initial masses similar to the Sun ($\sim0.8-1.3$\,\Msun), the relation between initial period and resulting WD mass will change only slightly, as the Roche-volume of the more massive star will also change only slightly. Therefore, our conclusion for the occurrence of tertiaries remains true as long as the initial masses are similar to that of the Sun. ELM WDs can also form from close binaries consisting of significantly more massive stars. In this case the non-degenerate star must be more massive as well and the triple statistic from \citet{Tokovinin06} no longer applies.  

\begin{figure}
    \includegraphics[width=\columnwidth]{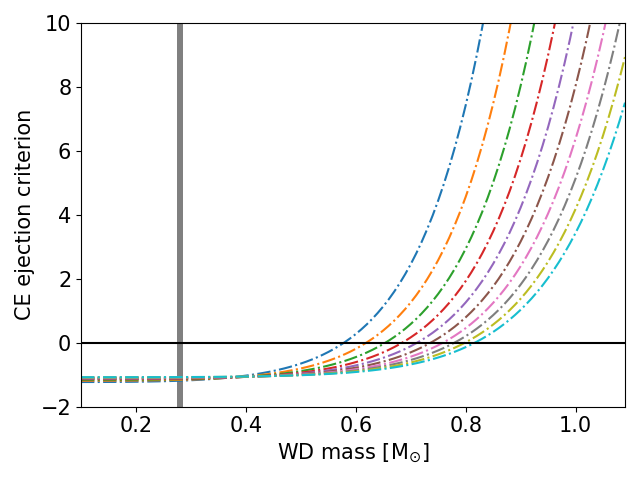}
    \caption{CE ejection criterion (left side of Equation \ref{eq:CE criterion}) using $\alpha=1$ and $\lambda=2$ as function of the WD mass for TYC\,6992. Dash-dotted lines consider initial mass ratios from 1 (blue) to 3 (cyan) in steps of 0.2. For the observed orbital configuration of TYC\,6992 CE evolution is only possible for WD masses $\gtrsim 0.56\,\mathrm{M_{\odot}}$ (i.e. above the black line at $y=0$). Therefore, a WD with $0.28\pm0.01\,\mathrm{M_{\odot}}$ (gray area) like the one in TYC\,6992 could not have formed via CE evolution.}
    \label{fig:CE criterion}
\end{figure}

\begin{figure}
    \centering
    \includegraphics[width=\columnwidth]{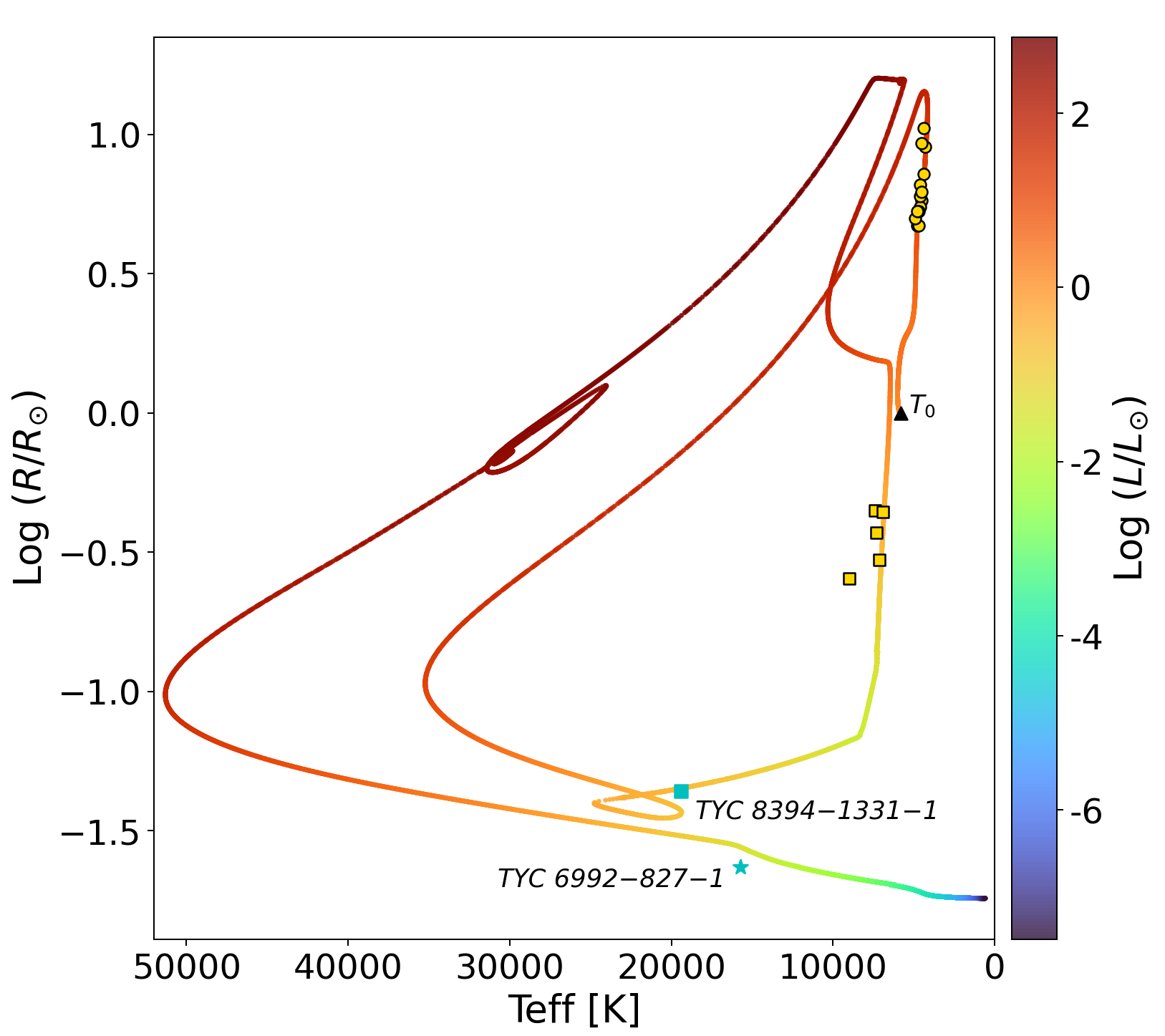}
    \caption{Radius evolution plotted against the effective temperature of a main-sequence binary with initial stellar masses of 1.1 and $0.95\,\mathrm{M_\odot}$ and an initial orbital period of 4.55\,d (T$_0$ is indicated with the black arrow which also show the direction of the evolution), evolved in MESA. The color code of the line corresponds to the luminosity evolution of the $1.1\,\mathrm{M_\odot}$ WD progenitor. Yellow dots and squares represent the WD progenitors from \citet{El-badry2022} and EL\,CVns from \citet{Lagos_2020} respectively, while the blue star represent TYC\,6992. The blue square show the spot that correspond to the similar binary TYC\,8394 from \citet{Parsons_2023}.}
    \label{fig:mesa-tyc}
\end{figure}

\begin{figure}
    \centering
    \includegraphics[width=\columnwidth]{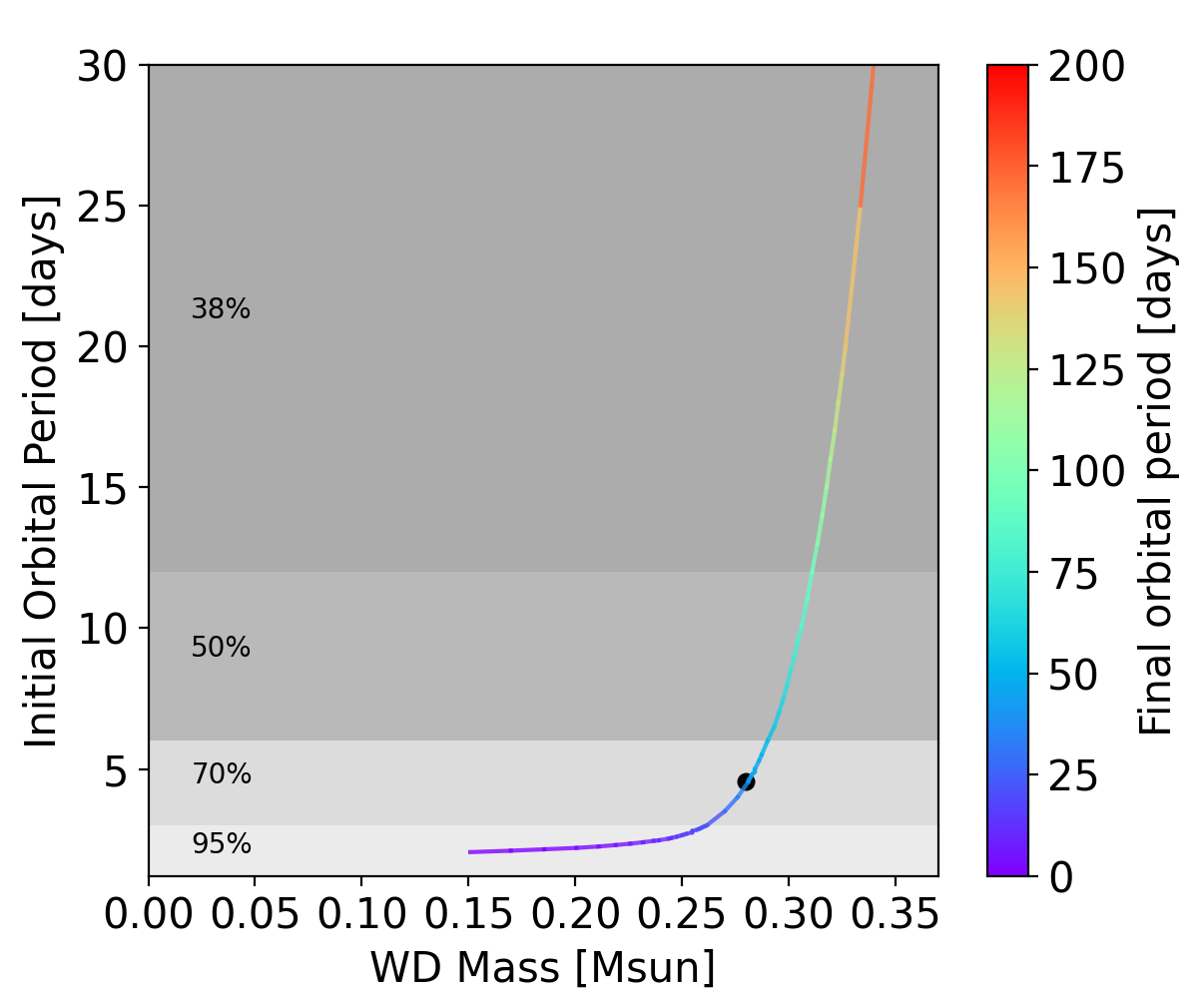}
    \caption{The mass of the helium core formed during the MESA evolution of a main-sequence star with a mass of $1.1\,\mathrm{M_\odot}$ is expressed in terms of the initial orbital period of the binary system. The black dot corresponds to observations of TYC\,6992-827-1, while the color code of the line corresponds to the final orbital period of each binary simulated with MESA. The grey shaded areas indicate the boundaries of the initial orbital period and the corresponding probability of the system having a tertiary companion \citep[taken from Fig. 14 of ][]{Tokovinin06}.}
    \label{fig:mesa-tokl}
\end{figure}

\section{The future of TYC 6992-827-1}
\label{sec:future}
\begin{figure}
    \centering
    \includegraphics[width=\columnwidth]{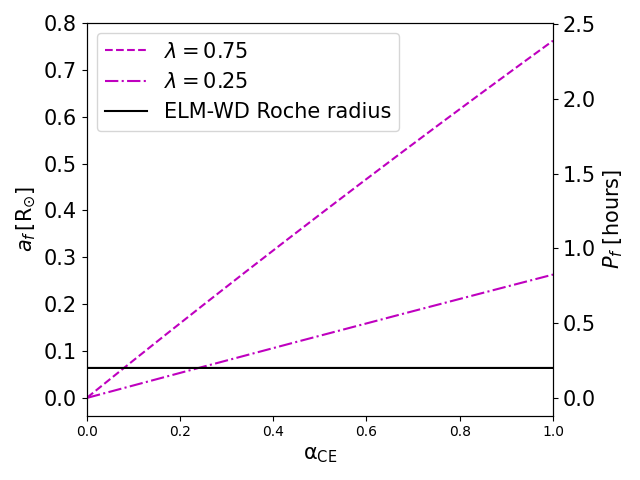}
    \caption{Final orbital semi-major axis (or its equivalent orbital period  $P_\mathrm{f}$) of the post CE double WD TYC\,6992 as function of the common envelope efficiency parameter $\alpha$ for two limiting values of $\lambda$. The solid black line represents the minimum semi-major axis before the ELM WD fills its Roche lobe. For highly  efficient envelope ejections (i.e. $\alpha \approx1$) the value of $a_\mathrm{f}$ ($P_\mathrm{f}$) ranges from 0.26 (0.82) to 0.76 (2.4) $\mathrm{R_{\odot}}$ (h), while for less efficient ejections ($0.08 \lesssim \alpha \lesssim 0.25$) the minimum value for $a_\mathrm{f}$ ($P_\mathrm{f}$) is $\approx 0.06 \mathrm{R_{\odot}}$ ($ \approx 0.2$\,h).}
    \label{fig:PCE_scenario}
\end{figure}

The detached phase of TYC\,6992, which might include a short symbiotic phase \citep{Parsons_2023},  will continue until the SG star fills its Roche lobe, starting a second mass transfer episode. Since the mass ratio of the binary is well above the critical mass ratio ($q_{\text{crit}} \approx 0.7-1$), mass transfer will proceed unstably, triggering a CE phase. If the envelope is ejected, then the orbital semi-major axis of the resulting double WD binary is given by

\begin{equation}
    a_{f}=\frac{M_\mathrm{SG,core}M_\mathrm{ELM}}{2M_\mathrm{SG}\left(\frac{M_\mathrm{ELM}}{2a_i} - \frac{M_\mathrm{SG,env}}{\lambda \alpha R_\mathrm{SG}} \right)},
\end{equation}

where $M_\mathrm{SG,core}$, $M_\mathrm{SG,env}$ and $M_\mathrm{SG}$ are the core, envelope and total mass of the SG star, respectively, $R_\mathrm{SG}$ the radius of the SG star at the onset of CE,  $M_\mathrm{ELM}$ the mass of the ELM WD, and $\mathrm{a_{i}}$ the semi-major axis at the onset of the CE phase. By taking a referential $1.3\,\mathrm{M_{\odot}}$ evolutionary track from the MESA Isochrones and Stellar Tracks  \citep[MIST, ][]{Paxton2011,Dotter2016,Choi2016}, we have $M_\mathrm{SG,core}=0.32\,\mathrm{M_{\odot}}$, $M_\mathrm{SG,env}=0.97\,\mathrm{M_{\odot}}$ and $R_\mathrm{SG}=30\,\mathrm{R_{\odot}}$. For envelope-structure values consistent with sub-giant stars \citep[$\lambda \approx 0.25-0.75$][]{Claeys_2014}, two outcomes are possible depending on the common envelope ejection efficiency (Fig. \ref{fig:PCE_scenario}): if $\alpha \gtrsim 0.25$, the envelope is ejected and the resulting double WD would have an orbital period between 3\,min and 3.6\,h (similar to the population of tight WD+WD binaries predicted by \citet{Shariat2023}, their figure 6). At such orbital periods, angular momentum loss due to gravitational radiation becomes important, leading to the merger of the two WDs on timescales ranging from approximately $0.3$\,Myr to a few Gyr \citep{Brown2016}. On the other hand, if $\alpha \lesssim 0.25$, the binary is prone to a merger event, likely ending up as a single WD somewhat more massive than the current ELM WD, as suggested by \citet{Parsons_2023}. This scenario is consistent with the merging population of ELM WD + red giant binaries with orbital periods $>10$\,d  predicted by \citet[][their Fig. 6]{Shariat2024}.

In both outcomes, the resulting wide binary composed of a WD and a K/M-type companion (the current tertiary) will resemble the configuration of HE\,0430–-2457, the first ELM WD found in a wide binary with a K-type companion, whose origin is also attribute to the merger of an inner binary in a hierarchical triple \citep{Vos_2018}. Given its similarity to TYC\,6992, both scenarios, the survival of a close binary inner binary or the merger of the inner binary, also apply to TYC\,8394.

\section{Discussion} 

We have shown that the current configuration of the two ELM WD + SG binaries discovered by \citet{Parsons_2023} can only be explained if the former main-sequence binary has an orbital period $\lesssim 5$\,d and passes through a phase of stable mass transfer when the ELM WD progenitor is at the end of (or just leaving) the main-sequence. The discovery of tertiary companions around these systems, either by direct imaging or radial velocity monitoring, provides further support to this formation channel and highlights their similarities to EL\,CVn-type binaries. Given that the main difference between both type of binaries is the evolutionary stage of the stellar components, ELM WD + SG binaries are most likely descendants of EL\,CVn binaries.   

The past history of EL\,CVn-like binaries has recently been unveiled by \citet{El-badry2022}. They discovered 14 binaries composed of a highly stripped low-mass giant (the future ELM WD) and a $\approx 2\, \mathrm{M_{\odot}}$ main-sequence companion. Using MESA simulations they found evolution tracks similar to those derived by \citet{Chen_2017} for EL\,CVn binaries. Additionally, they identified similar main-sequence binary progenitors: low-mass giant progenitors with $\approx 1\,\mathrm{M_{\odot}}$, initial mass ratios  $0.6 \lesssim q_\mathrm{i} \lesssim 0.9$ and orbital periods of $\approx 1$\,d \citep[see figure 4 of ][]{Chen_2017}. This sample of pre-EL\,CVn binaries also perfectly aligns with the stellar radius-effective temperature evolutionary track of our canonical binary setup 
shown in Fig. \ref{fig:mesa-tyc}. Hence, it is reasonable to expect that these pre-EL\,CVn systems should also have tertiary companions. In an attempt to look for them, we cross-match the sample of pre-EL\,CVn binaries with the Gaia-based catalog of resolved binaries of \citet{El-Badry_2021_wide_binaries} and find that the closest ($\approx 720$\,pc) pre-EL\,CVn system (Gaia DR3 5243109471519822720) has in fact a wide companion (Gaia DR3 5243109471512749056) with a projected separation of $\approx 2375$\,au. Since the remaining 13 systems have distances $\gtrsim 900$\,pc, the identification of additional companions is severely hindered by the angular resolution (and sensitivity) of Gaia. To improve the likelihood of detection, high-angular resolution and high-contrast imaging are required \citep[e.g. ][]{Lagos_2020}.

It is worth noting that \citet{Garbutt_2024} and \citet{Shahaf_2024} identified a sample of WDs with main sequence companions of spectral type F, G, K with long periods (exceeding several hundred days) that are not consistent with the predictions of stable mass transfer (see their figure\,8). Systems with WD masses exceeding $0.6$\,\Msun\ are most likely post common envelope binaries that started mass transfer when the WD progenitor was on the late asymptotic giant branch \citep{Belloni_a2024,Belloni_b2024}. 
However, those binaries with orbital periods of hundreds of days but containing ELM WDs cannot have formed through common envelope evolution and, according to the orbital period-WD mass relation shown in the upper panel of Fig.\,8 in \citet{Garbutt_2024}, are also difficult to explain through stable mass transfer. A dedicated survey for tertiaries to these close binaries could provide evidence for their formation through stable mass transfer. If the existence of tertiaries can be confirmed at a high rate ($\gtrsim70$ per cent), a currently unknown mechanism producing wider orbits than predicted by standard models of stable mass transfer must be at work. 

A similar puzzling situation has been claimed to exist for the self-lensing binary KIC 8145411, initially identified as a $0.2\,\mathrm{M_{\odot}}$ ELM WD with an orbital period of $456$\,d around a solar-type companion \citep{Masuda_2019}. However, \citet{Yamaguchi_2024} discovered an additional unresolved tertiary companion that altered the mass estimation, revealing that the WD mass is larger than previously thought ($0.53\,\mathrm{M_{\odot}}$) and consistent with CE evolution. 
In principle, the presence of additional unresolved companions to the astrometric ELM WD + MS binaries could similarly alter the estimation of the corresponding WD masses. However, again, this potential solution remains speculative unless more unresolved tertiaries are discovered.

The fact that most ELM WD with non-degenerate companions are inner binaries of hierarchical triples raises the question of whether triple dynamics is primarily responsible for their formation by triggering the inward orbital migration necessary for stable mass transfer. While \citet{Fabrycky_Tremaine2007} showed that KCTF is able to produce close (P$<10$\,d) binaries, \citet{Moe_Kratter2018} found that KCTF is only able to produce $\approx40$ per cent of such binaries. The remaining $60$ per cent are most likely formed due to disc fragmentation followed by inward migration caused by energy dissipation from interactions with the primordial gas during the pre-main-sequence. As discussed by \citet{Lagos_2020}, the presence of tertiary companions around EL\,CVn binaries might be a fingerprint of the initial formation conditions. Either the initial mass required for migration to work is large enough that it is virtually always accompanied by the formation of a tertiary via core fragmentation or the tertiary is triggering disc fragmentation at an early stage when there is still enough gas for migration to take place.

For the five EL\,CVn binaries found by \citet{Lagos_2020} and the pre-EL\,CVn in the \citet{El-Badry_2021_wide_binaries} catalogue, we find that the ZLK oscillation timescale is on the order of several Gyr in most cases, supporting the pre-main-sequence migration scenario. This estimate is derived by using half of the projected separation of the tertiary (to account for potential orbital widening due to mass loss) as a proxy of the outer semi-major axis, while assuming an outer eccentricity of $0.7$, inner period of five days and nominal stellar masses for EL\,CVn systems (i.e. a $1 \mathrm{M_{\odot}}$ twin inner binary and $0.8 \mathrm{M_{\odot}}$ tertiary). The pre-main-sequence migration scenario is also consistent with the twin and close nature of the EL\,CVn progenitor binaries. As discussed by \citet{Tokovinin2000}, the population of twin solar-type close binaries can be explained by primordial circumbinary discs massive enough to simultaneously enable accretion and inward migration, while also enhancing the possibility of capturing a tertiary companion by dissipating its gravitational energy.

While the smaller projected separations of the tertiary companion in TYC\,6992 and the highly compact nature of TYC\,8394 (with an outer semi-major axis of $\approx 2.2$\,au) suggests that KCTF could have played a role in the formation of the triple configuration prior to mass transfer, this scenario is somewhat disfavored by the fact that triple systems with outer projected separations $\lesssim50$\,au tend to have coplanar orbits \citep{Tokovinin2017}. The discovery and characterization of additional close tertiary companions around ELM WDs with non-degenerate companions could offer further insights into their formation mechanism(s).



\section{Conclusions}
We report the discovery, through direct imaging, of a tertiary companion around the binary TYC\,6992-827-1, consisting of a sub-giant star and a ELM WD. By reconstructing its evolutionary history, we found that it  follows the same evolutionary path as EL\,CVn binaries, which are composed of a pre-ELM WD and a main-sequence companion of spectral type A/F. Our binary evolution simulations also naturally reproduce the observed stellar radius-effective temperature relation of the ELM WD progenitor star in pre-EL\,CVn binaries. Therefore, TYC\,6992-827-1 (and its sibling TYC 8394-1331-1) represent an later evolutionary stage of EL\,CVn binaries in which the companion to the ELM WD just left the main sequence. This is consistent with the triple nature of TYC\,6992-827-1 and TYC 8394-1331-1 and most EL\,CVn binaries \citep{Lagos_2020} which indicates that all these systems form from short period main sequence binary stars. EL\,CVn binaries and their late evolutionary stage in the form of ELM WD + sub-giant binaries are currently the only known examples where an ELM WD has a non-degenerate companion.

We estimate that TYC\,6992-827-1 (and also TYC 8394-1331-1) will experience a common envelope phase in the future. Depending on how efficient orbital energy is used to expel the envelope, the final product could be either a single WD or a tight double WD binary which would merge into a single WD within a few Gyr. Thus, the former triples system will become wide binaries composed of a WD and a K-type main-sequence companion.

\section*{Acknowledgements}
We thank the anonymous referee for their helpful comments and suggestions.
This project has received funding from the European Research Council (ERC) under the European Union’s Horizon 2020 Research and Innovation Programme (grant agreement no. 101020057). MRS thanks for support from FONDECYT (grant number 1221059). 
Based on observations collected at the European Southern Observatory under ESO programme 110.243S.001.

\section*{Data Availability}

The data and numerical tools used in this article can be obtained
upon request to the corresponding author and after agreeing to the
terms of use.



\bibliographystyle{mnras}
\bibliography{mnras_template} 








\bsp	
\label{lastpage}
\end{document}